\documentclass[12pt]{iopart}

\usepackage{geometry}
\usepackage{xcolor}
\usepackage{rotating}
\usepackage[utf8]{inputenc}
\usepackage[T1]{fontenc}
\usepackage[hyphens]{url}
\usepackage[colorlinks,pdftex,pdfpagelabels,bookmarks,hyperindex,hyperfigures]{hyperref}
\hypersetup{
  pdftitle={},
  pdfauthor={Reza Sameni},
    pdfnewwindow=true,      
    colorlinks=true,       
    linkcolor=blue,          
    citecolor=magenta,        
    filecolor=cyan,      
    urlcolor=teal
}

\usepackage{color}
\usepackage{amssymb}
\usepackage{graphicx}
\usepackage{subcaption}
\usepackage{algorithm}
\usepackage{algorithmic}
\usepackage{balance}

\begin{document}



\title{ECG-Image-Database: A Dataset of ECG Images with Real-World Imaging and Scanning Artifacts; A Foundation for Computerized ECG Image Digitization and Analysis}

\author{Matthew~A.~Reyna\textsuperscript{1}, Deepanshi\textsuperscript{1}, James~Weigle\textsuperscript{1}, Zuzana~Koscova\textsuperscript{1}, Kiersten~Campbell\textsuperscript{1}, Kshama~Kodthalu~Shivashankara\textsuperscript{2}, Soheil Saghafi\textsuperscript{1}, Sepideh~Nikookar\textsuperscript{1}, Mohsen~Motie-Shirazi\textsuperscript{1}, Yashar~Kiarashi\textsuperscript{1}, Salman~Seyedi\textsuperscript{1}, Gari~D.~Clifford\textsuperscript{1,3}, and Reza~Sameni\textsuperscript{1,3}}
\address{\textsuperscript{1}Department of Biomedical Informatics, Emory University, Atlanta, GA 30322, USA}
\address{\textsuperscript{2}School of Electrical and Computer Engineering, Georgia Institute of Technology, Atlanta, GA 30332, USA}
\address{\textsuperscript{3}Department of Biomedical Engineering, Georgia Institute of Technology, Atlanta, GA 30332, USA}
\ead{\href{rsameni@dbmi.emory.edu}{rsameni@dbmi.emory.edu}}

\markboth 
{ECG-Image-Database: An Open-Access ECG Image Dataset}
{ECG-Image-Database: An Open-Access ECG Image Dataset}




\vspace{10pt}
\begin{indented}
\item[]September 2024
\end{indented}

\begin{abstract}
\textit{Objective:} We introduce ECG-Image-Database, a large and diverse collection of electrocardiogram (ECG) images generated from ECG time-series data that exhibit real-world scanning, imaging, and physical artifacts. This dataset addresses the need for digitizing paper-based and non-digital ECGs for computerized analysis, providing a foundation for developing robust machine and deep learning models capable of ECG image digitization.

\textit{Approach:} We used ECG-Image-Kit, an open-source Python toolkit, to generate realistic images of 12-lead ECG printouts from raw ECG time-series data. We include images with realistic distortions, including noise, wrinkles, stains, and perspective shifts generated both digitally and physically.
The toolkit was applied to 977 12-lead ECG records from the PTB-XL database and 1,000 from Emory Healthcare to create high-fidelity synthetic ECG images. These 1,977 unique images were subjected to both programmatic distortions using ECG-Image-Kit and to physical effects, such as soaking, staining, and mold growth, followed by scanning and photography under various lighting conditions to create real-world artifacts, producing a total of 35,595 images.

\textit{Main results:} The resulting dataset includes 35,595 software-labeled ECG images (generated from 1,977 unique records) from different sources (in Germany and the USA) with a wide range of imaging artifacts and distortions. The dataset provides ground truth time-series data alongside the corresponding images, offering a reference for training and evaluating machine and deep learning models for ECG digitization and classification. The images vary in quality, ranging from clear scans of clean papers to noisy photographs of degraded papers, enabling the development of more generalizable and robust digitization algorithms.

\textit{Significance:} ECG-Image-Database addresses a critical gap in cardiovascular research by supporting the development of machine learning models capable of classifying or converting ECG images into time-series data, ensuring that valuable diagnostic information from non-digital archives is not lost. ECG-Image-Database aims to serve as a reference for ECG digitization efforts and opens new avenues for research in computerized ECG analysis and diagnosis. The dataset was used in the PhysioNet Challenge 2024 on ECG image digitization and classification.
\end{abstract}



\maketitle

\section{Introduction}
Cardiovascular diseases (CVDs) are the leading cause of death globally. To date, the electrocardiogram (ECG) remains the most accessible tool for CVD assessment, with over three million ECGs generated daily by clinicians \cite{shenasa2015ecg}, and millions more each day from wearable and personal devices. It is estimated that there are billions of digital diagnostic ECGs globally \cite{tison2019}, with an equal number in hardcopy formats, including microfilm,  paper, and scanned images \cite{davis1973electrocardiogram}. Paper ECGs are gradually being replaced by digital ECGs (although much more slowly in low-income and low-resource regions), but legacy ECG records contain invaluable information on individual history, rare events, and the evolution of CVDs across generations and geography. Moreover, the proprietary nature of many ECG acquisition systems means that sharing the raw ECG is often difficult or costly, and printing out the ECG as a PDF, image, or on physical paper is frequently the only option for moving the data beyond the commercial walled gardens. Due to natural deterioration and a lack of funding for physical archives, decades of non-digital ECG archives worldwide will soon be destroyed before we can learn from them and leverage them for training machine learning (ML) models for algorithmic diagnosis of CVDs. This loss would be irreversible and an almost incalculable setback for cardiovascular research, as the ECG is the only biological signal that has been recorded for over a century without significant changes in its acquisition protocol. Over recent decades, most countries have experienced significant shifts in cardiovascular health, and longitudinal analysis of paper records has the potential for us to map these changes in an unprecedented manner. Moreover, the populations from which we can perhaps learn the most and which are least represented in commercial products are the most likely to have paper ECGs. 

Modern ECG machines and wearable technologies enable the collection of massive digital ECG datasets. However, it will take decades for digital archives to begin replacing the lost diagnostic and demographic information from physical archives, especially in low-resource areas and low- and middle-income countries (LMICs), where electronic health records are rare, leading to the loss of demography-specific clinical data. Even in high-income regions, printed ECGs remain popular and are routinely circulated between experts as hardcopies or pictures for adjudication and training purposes. Recently, specialized public forums have formed on social media platforms
, with hundreds of thousands of expert members who regularly share de-identified (protected health information redacted) ECG images and discuss diagnoses and clinical outcomes, serving as a new form of the traditional \textit{grand rounds} or \textit{case studies} in medical education and inpatient care.

To preserve printed ECGs, some research teams and healthcare systems in affluent regions scan and archive them in image formats. However, there are insufficient incentives to dedicate the time and effort to preserving most printed ECGs.
Scanning alone is not enough to ensure the usefulness of an ECG because: i) ECG images are incompatible with state-of-the-art computerized ECG annotation software, which are typically trained on and analyze ECG time-series; ii) ECG images are not currently automatically searchable for annotations and anomalies; iii) existing technologies for ECG diagnosis are not available at the point-of-care (POC); and iv) it is hard to determine \emph{a priori} if the scanned ECGs are of acceptable quality for use in training ML models for ECG diagnosis. To address these issues, research teams have developed image processing pipelines, supervised computer vision software, and more recently, deep learning-based methods for ECG digitization. The highlights of these contributions are reviewed in Section \ref{sec:background}.

The ECG digitization process can be understood as a sequence of steps: i) scanning or photographing physical ECGs as images, ii) extracting ECG time series data from the images, and iii) annotating or labeling the ECG data. This process works better when the later steps inform the earlier ones; for example, a device automatically rescans a paper ECG when the image is not clear enough for time-series extraction. However, to date, paper-based ECG digitization and algorithmic ECG diagnosis have not yet been integrated using domain knowledge of the ECG. Therefore, the added value of learning from historical ECG images remains unrealized.


While there is an increasing availability of ECG time-series datasets, there remains a significant gap in the availability of standardized datasets that pair these time-series with corresponding ECG images across a range of qualities. Currently, no standard datasets provide the underlying ECG time-series alongside ECG images that vary from high-quality scans to photographs, low-quality grayscale or black-and-white versions, and those affected by environmental artifacts such as imaging and scanning issues, natural deterioration, and other common distortions. To address this critical gap in ECG image digitization, we have developed tools to convert ECG time-series data from standard, widely accessible datasets, including the PTB-XL dataset \cite{wagner2020ptb,PTB-XL-PhysioNet}, as well as ECG data from a demographically diverse population at Emory Healthcare in the USA \cite{Reyna2021}, into images that replicate the form familiar in medical settings. These images were then printed, contaminated with natural artifacts, and compiled into a very large dataset that encompasses a wide spectrum of imaging and scanning artifacts. This comprehensive dataset enables the development and training of more robust and generalizable digitization models, ensuring that the tools can effectively handle the diverse conditions encountered with real ECG images in real-world clinical settings.

Importantly, although evident in the signal and image processing communities, there is a significant difference between ECG data in time-series format and images of ECG signals. Throughout this paper, ``ECG time-series'' refers to the digital samples of an ECG, i.e., a sequence of ``signal'' amplitude values over time, which is the preferred format for ECG analysis and annotation software; ``non-digital ECG'' refers to ECGs that are in a physical hardcopy format (such as paper or microfilm) and have not been digitized; ``scanned ECG'' or ``ECG image'' refers to pictures of ECGs, obtained either through photography or scanning of non-digital ECGs, or stored in an image format directly by an ECG machine at the point-of-care. Although the latter is considered a digital representation of the ECG, their corresponding ECG time series are generally not directly available (unless stored as a PDF with the raw ECG data embedded within it, which is somewhat rare in our experience). Throughout this article, ``ECG digitization'' refers to the process of retrieving ECG time-series data from images, and ``ECG annotation'' refers to the process of assigning diagnostic labels to an ECG (in image or time-series formats), either by a human or a machine/algorithm.

\section{Background and significance}
\label{sec:background} 

To provide context, we review some of the major highlights of previous research on ECG image digitization and clinical measurements, which necessitate the development of standardized ECG image datasets.

Traditional image processing and computer vision pipelines focus on digitizing ECG data by removing background grids and noise to extract clear signals. However, these approaches differ in their techniques and algorithms. Common methods include image processing techniques such as binary morphological image operations \cite{Tun2017}, grayscale thresholding \cite{Ravichandran2013}, and linear filtering \cite{Ganesh2021}. Additionally, several studies have used optical character recognition (OCR) to capture patient demographic information and integrate it into medical records \cite{Ravichandran2013, Ganesh2021, Silva2008}. Some methods have also considered cost-effective, non-hardware solutions \cite{Silva2008, virgin2018conversion}.

Fortune et al.\ in \cite{Fortune2022} developed an algorithm to capture ECG morphology and beat timing with high precision, validated through statistical measures. In \cite{Zhang1987}, image processing techniques such as histogram filtering were applied to remove noise, allowing for efficient archival storage. Widman et al.\ in \cite{Widman1991} employed optical scanners, demonstrating improved signal fidelity with adjustments in paper speed and amplifier gain. Reproducible methods for scanning and analyzing QT intervals were introduced in \cite{Bhullar1993} and \cite{Wang1996}, enhancing the clinical utility of ECG digitization techniques. Lobodzinski et al.\ \cite{Lobodzinski2002} developed an optical ECG waveform recognition method to digitize paper ECGs using statistical filtering and image processing. This was further enhanced in \cite{Lobodzinski2003}, where the authors proposed an XML-based format for storing the digitized data. Badilini et al.\ in \cite{Badilini2005} introduced ECGScan, employing active contour modeling to detect and extract ECG waveforms from paper records. Mitra et al.\ in \cite{Mitra2004} applied Fourier transform techniques to convert ECG images into digital signals, focusing on frequency analysis. Karsikas et al.\ in \cite{Karsikas2007} used a digitization method involving scanning paper ECGs and applying image processing to correct misalignment and remove grid noise, preserving ECG signal integrity.

More recently, advanced machine learning pipelines have been proposed for converting ECG images into time-series data. As demonstrated by Helkeri et al.\ \cite{Holkeri2018}, traditional machine learning techniques can be leveraged to refine and improve the accuracy of clinical measurements from ECG images. Baydoun et al.\ in \cite{Baydoun2019} applied neural networks to convert ECG scans into actionable data.

However, these approaches have largely been constrained by the availability of small, manufacturer-specific datasets, limiting their broader applicability and generalizability. Despite these advances, the full potential of deep learning models such as ResNet \cite{He2016} and large-scale datasets like ImageNet \cite{Deng2009}, which have significantly advanced image processing applications in other domains, remains underexplored in the context of ECG digitization. The adoption of these more powerful models could lead to more robust and scalable solutions for ECG image processing, particularly by addressing the limitations of small datasets and manufacturer-specific variations. These generic computer vision tools have yet to address the unique challenges of ECG digitization \cite{Waits2017}, including the diversity of diagnostic ECG devices ranging from single to twelve leads, the diversity of ECG paper standards, creases, wrinkles, non-uniform fading of ink, and other physical wear, as well as the extraction of computerized or handwritten text such as patient identifiers, lead names, and notes, and low-resolution images or scans. In \cite{Waits2017}, the challenges of digitization approaches prior to the last decade are categorized into several key issues: grid-related problems, time efficiency, alignment difficulties, signal quality and noise concerns, and validation challenges, particularly regarding validation on large and diverse samples. More recent approaches, like deep learning models, have the potential to learn ECG-specific features directly from images without the need to extract the underlying ECG time-series data \cite{Brisk2019}. which used a deep neural network to classify down-sampled images produced from the PhysioNet Challenge 2017 \cite{PhysioNetChallenge2017} dataset's raw training signals. However, the quality metrics of generic computer vision deep models differ from those needed for ECG diagnosis, and the volume of available ECG images with simultaneous ECG time-series and clinical annotations is currently inadequate for training deep learning models. Data augmentation with synthetic ECG images (time-series printed on paper ECG grids) may allow algorithms to diagnose ECG images directly. More recently, \cite{Shivashankara2024} introduced deep neural networks to enhance the ECG digitization process, demonstrating promising results in handling variability in ECG images across devices using data augmentation with synthetic paper ECG images. Similarly, \cite{PMcardio-2024} introduced a two-stage, deep-learning based approach for ECG digitization and classification across a variety of image types, sources, and variation.

\section{Generating realistic ECG images from time-series data}
ECG-Image-Kit is an open-source toolbox developed to generate synthetic ECG images that closely replicate real-world ECG printouts \cite{Shivashankara2024,ecg-image-kit-software}. The primary purpose of this toolkit is to support the training and development of data intensive deep learning models for ECG digitization, addressing the critical need to convert legacy, non-digital ECG archives into digital formats compatible with modern diagnostic AI-ML tools. This tool was used to create the initial electronic version of the ECG-Image-Database presented in this work.

The toolkit creates realistic ECG images from time-series data, simulating the appearance of ECGs traditionally printed on thermal paper, or by modern inkjet and laser printers. To achieve this, ECG-Image-Kit introduces various customizable distortions that mimic the challenges encountered in real-world ECG digitization. The tool is able to overlay synthetic ECG images with both printed and handwritten text artifacts, including lead names, calibration pulses, patient information, and diagnostic notes. These artifacts are crucial in replicating the complexity of real paper ECGs, where such text often overlaps with the ECG waveform, making digitization a challenging task.

In addition to text artifacts, ECG-Image-Kit simulates physical distortions like wrinkles and creases, which are common in aged paper ECGs. These distortions are generated using advanced image processing techniques, such as image quilting and blurring, to ensure a realistic appearance. Furthermore, the toolkit can apply perspective transformations to the images, replicating the effects of non-standard camera angles and scanning imperfections often seen in ECGs captured via smartphones or low-quality scanners.

The toolkit also incorporates various imaging artifacts and noise, such as Gaussian noise, Poisson noise, and salt-and-pepper noise, to mimic different imaging conditions. Moreover, it can adjust the color temperature of the images to simulate the aging and environmental effects on ECG thermal paper. This approach to simulating real-world conditions ensures that the synthetic images generated by ECG-Image-Kit are as realistic as possible.

ECG-Image-Kit also offers flexibility in customizing the format of the ECG leads displayed in the synthetic images, supporting both standard and non-standard configurations. This feature is particularly important for replicating the diverse range of ECG formats encountered in clinical practice in different regions of the world. For example, standard paper ECGs typically display all 12 leads in 2.5\,s segments across four rows (sweeping from left to right in time). Additionally, leads II, V1, V2, and/or V5 are often shown as a continuous 10-second strip at the bottom for rhythm analysis. Older ECG machines recorded the 2.5\,s segments for different leads asynchronously, meaning the segments did not correspond to the same time frame. This is an important consideration for ECG digitization algorithms, as they cannot rely on the synchrony of channel segments to enhance the extracted ECG time series through multichannel post-processing. However, the longer strips do overlap with the shorter segments. The default mode of ECG-Image-Kit generates the most common 12 lead format (3 rows by 4 columns of 2.5\,s, with a long 10\,s strip on the bottom of the page). The default behavior can be changed through optional parameters detailed in \cite{ecg-image-kit-software}.

Additionally, the toolkit allows for the generation of large batches of ECG images with random, yet controlled, levels of artifacts and noise, making it ideal for creating very large datasets necessary for training deep learning models.

ECG-Image-Kit has been implemented in Python and was used in the George B. Moody PhysioNet Challenge 2024: Digitization and Classification of ECG Images to generate realistic ECG images with natural artifacts \cite{2024Challenge}. The toolkit provides specific command-line flags that allow users to control various aspects of generating ECG images from ECG time-series data stored in WFDB format \cite{WFDB-moody-software}. For instance, users can adjust the image resolution (with the \verb|-r| flag), add handwritten text distortions (using the \verb|--hw_text| flag), or introduce paper-like wrinkles and creases (with the \verb|--wrinkles| flag). The toolkit can process either individual ECG records or entire directories of ECG data. By specifying the path to the input directory or file and the desired output directory, users can generate ECG images in batch mode or for single records. This flexibility allows for granular control over the number and characteristics of the generated images, making it possible to create diverse representations of ECG data with varying levels of artifacts and distortions.

\section{ECG time-series data}
To create a rich dataset of ECG images that include various printing, scanning, and imaging artifacts, while having access to the ground-truth ECG as a time-series, two time-series datasets with clinical annotations were selected for this purpose. We used rejection sampling to extract a representative subset of 977 ECGs from the PTB-XL dataset and 1000 ECGs from the Emory dataset. These ECGs approximately preserved the univariate distributions of the patient attributes and classes from their source datasets. These datasets are described in the sequel.

\subsection{The PTB-XL dataset}

The PTB-XL dataset consists of 21,799 clinical 12-lead ECG records, each 10\,s long, from 18,869 patients \cite{wagner2020ptb}. These data were collected from Schiller AG devices between between October 1989 and June 1996. The patient group is 52\% male and 48\% female, ranging in age from 0 to 95 years (median age: 62). The dataset includes a broad range of heart conditions and healthy control samples, categorized into five main diagnostic classes: Normal ECG (9,514 records), Myocardial Infarction (5,469), ST/T Change (5,235), Conduction Disturbance (4,898), and Hypertrophy (2,649). The ECG data is stored in WFDB format at 500\,Hz and downsampled versions at 100\,Hz for convenience. Metadata for each record is found in a CSV file, including identifiers, demographic details, ECG diagnostics, and signal quality information. Additional fields track annotations such as heart axis, noise, and artifacts. The dataset includes a recommended 10-fold train-test split, with records in folds 9 and 10 having undergone human validation for label quality. PTB-XL+ \cite{Strodthoff2023,PTB-XL+PhysioNet} is a supplementary dataset that provides additional ECG features and algorithmic annotations from the PTB-XL dataset using two commercial algorithms (University of Glasgow ECG Analysis Program version R30.4.2 \cite{Macfarlane2005} and GE Healthcare's Marquette\texttrademark{} 12SL\texttrademark{} \cite{GE-Healthcare-Marquette}) and one open-source tool (ECGDeli version 1.1 \cite{Pilia2021}). The annotations include median beats, fiducial points, and automatic diagnostic statements, allowing users to train and evaluate machine learning models with the enhanced ECG metadata.

\subsection{The Emory Healthcare dataset}
We used a subset of 1,000 ECG records from the Emory Healthcare dataset, which features a diverse patient population in Georgia, USA. Access to this dataset was approved by Emory University's Internal Review Board under the PhysioCrowd protocol STUDY-00007353.

The Emory dataset consists of 12-lead clinical ECGs recorded between 2010 and 2022, by GE ECG machines of different generations, from healthcare subjects of different demographic background. The data were originally in XML format, containing ECG-based measurements and algorithmic annotations performed by GE's software. Due to the extended data collection period, we re-annotated the ECG time-series data using GE Healthcare's latest Marquette\texttrademark{} 12SL\texttrademark{} software, resulting in a unified relabeling of all the selected records.



\section{Dataset preparation}
\begin{figure}[htbp]
    \centering
    \rotatebox{90}{
        \begin{minipage}{\textheight}
            \centering
            \includegraphics[height=.35\columnwidth]{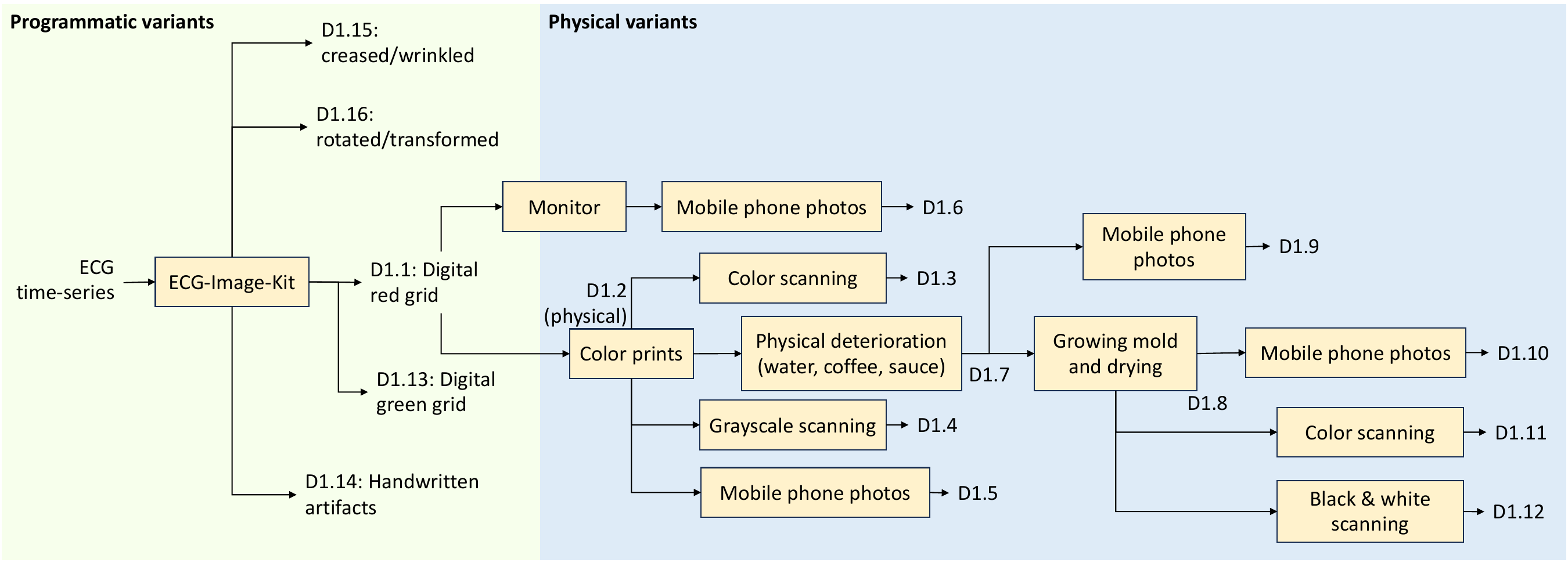}
            \caption{An illustration of the ECG-Image-Database creation procedure. The procedure has been repeated for 977 records of the PTB-XL dataset; a similar procedure applies to the 1000 diagnosis-stratified 12-lead ECG records from Emory Healthcare.}
            \label{fig:ecg-image-dat-diagram}
        \end{minipage}
    }
\end{figure}

\begin{figure}
\begin{subfigure}[t]{0.325\linewidth}
\includegraphics[width=\linewidth]{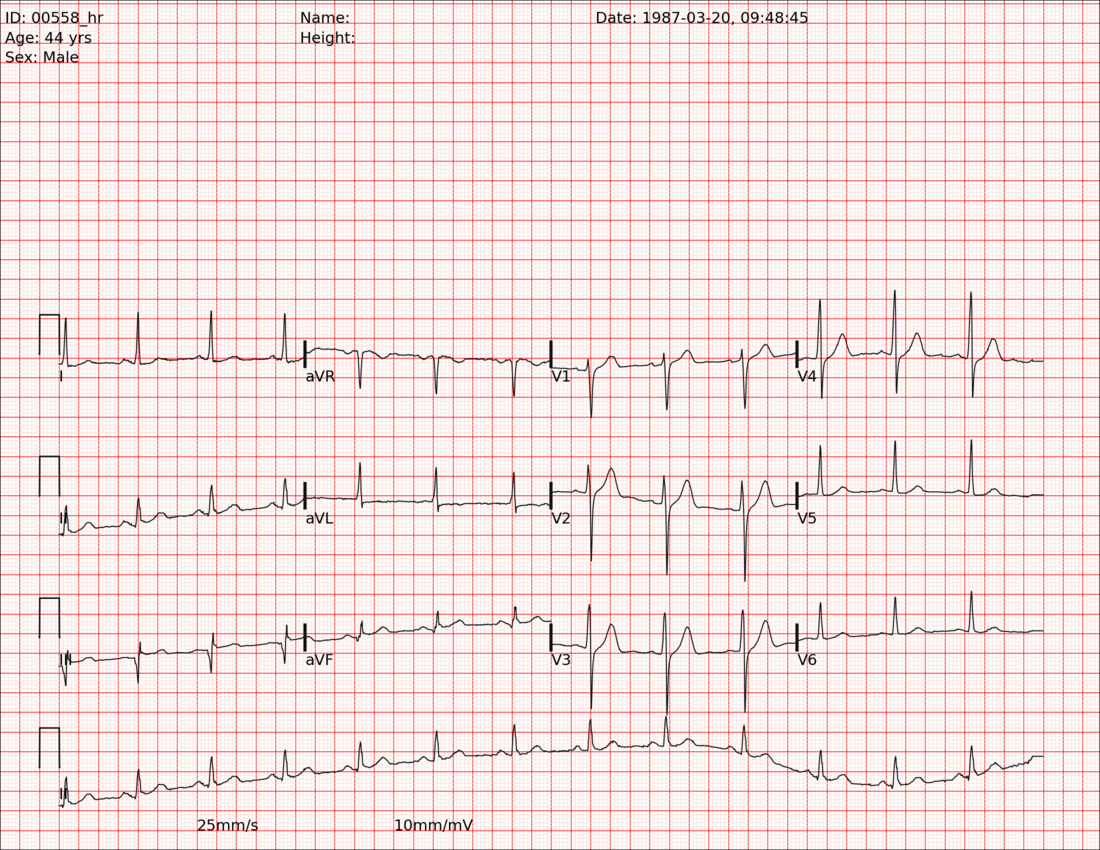}
\caption{Digital, red grid (\ref{sec:images_with_red_grid})}
\end{subfigure}
\hfill
\begin{subfigure}[t]{0.325\linewidth}
\includegraphics[width=\linewidth]{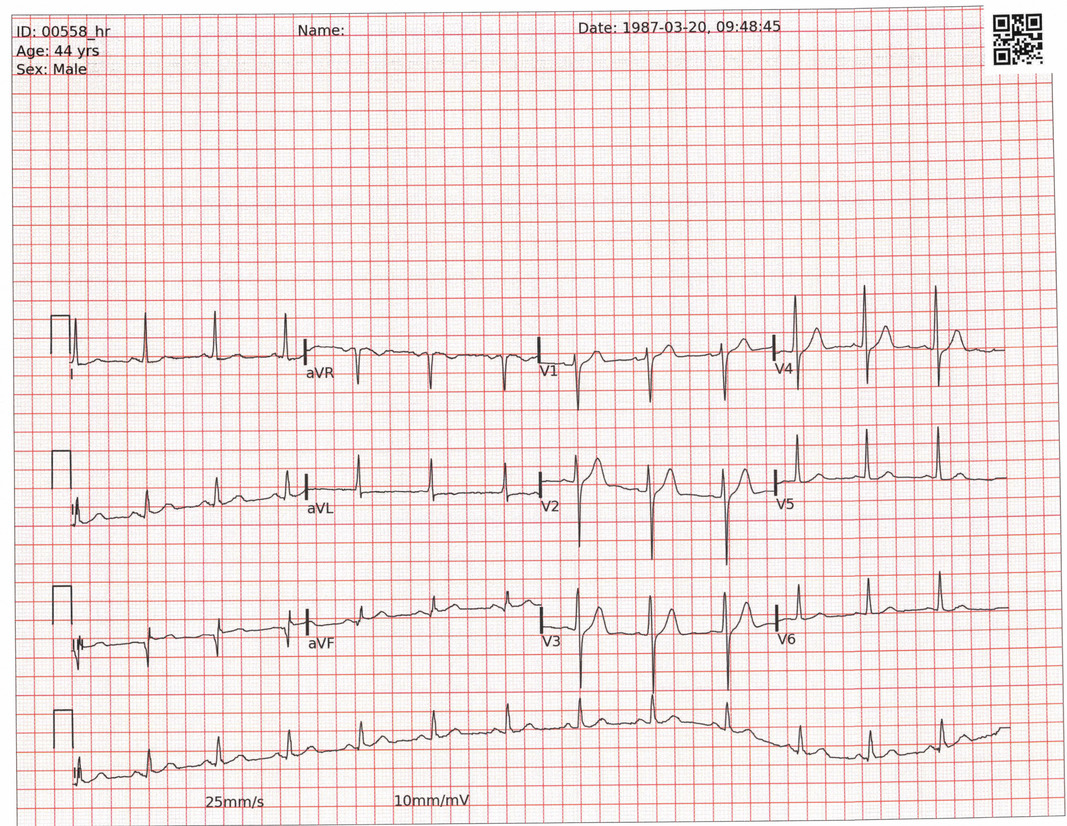}
\caption{Color scan of clean paper (\ref{sec:color-scan-clean-paper})}
\end{subfigure}
\hfill
\begin{subfigure}[t]{0.325\linewidth}
\includegraphics[width=\linewidth]{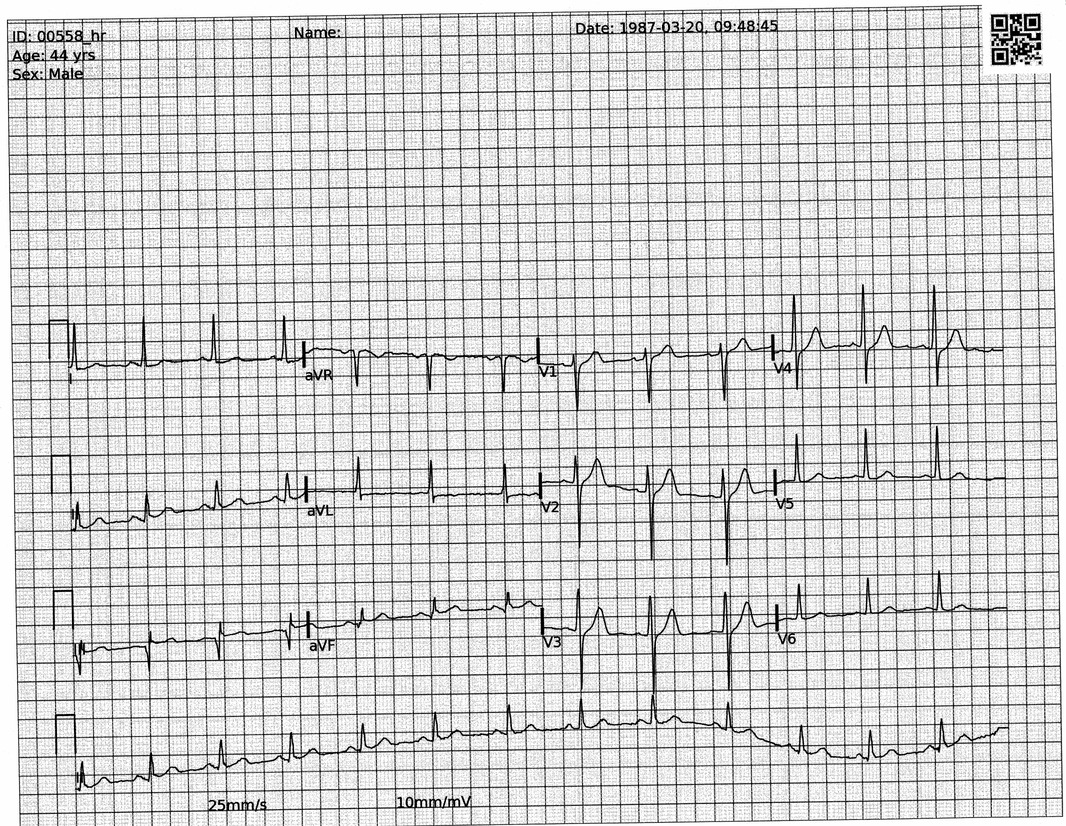}
\caption{Grayscale scan of clean paper  (\ref{sec:bw-scan-clean-paper})}
\end{subfigure}
\\
\begin{subfigure}[t]{0.325\linewidth}
\includegraphics[width=\linewidth]{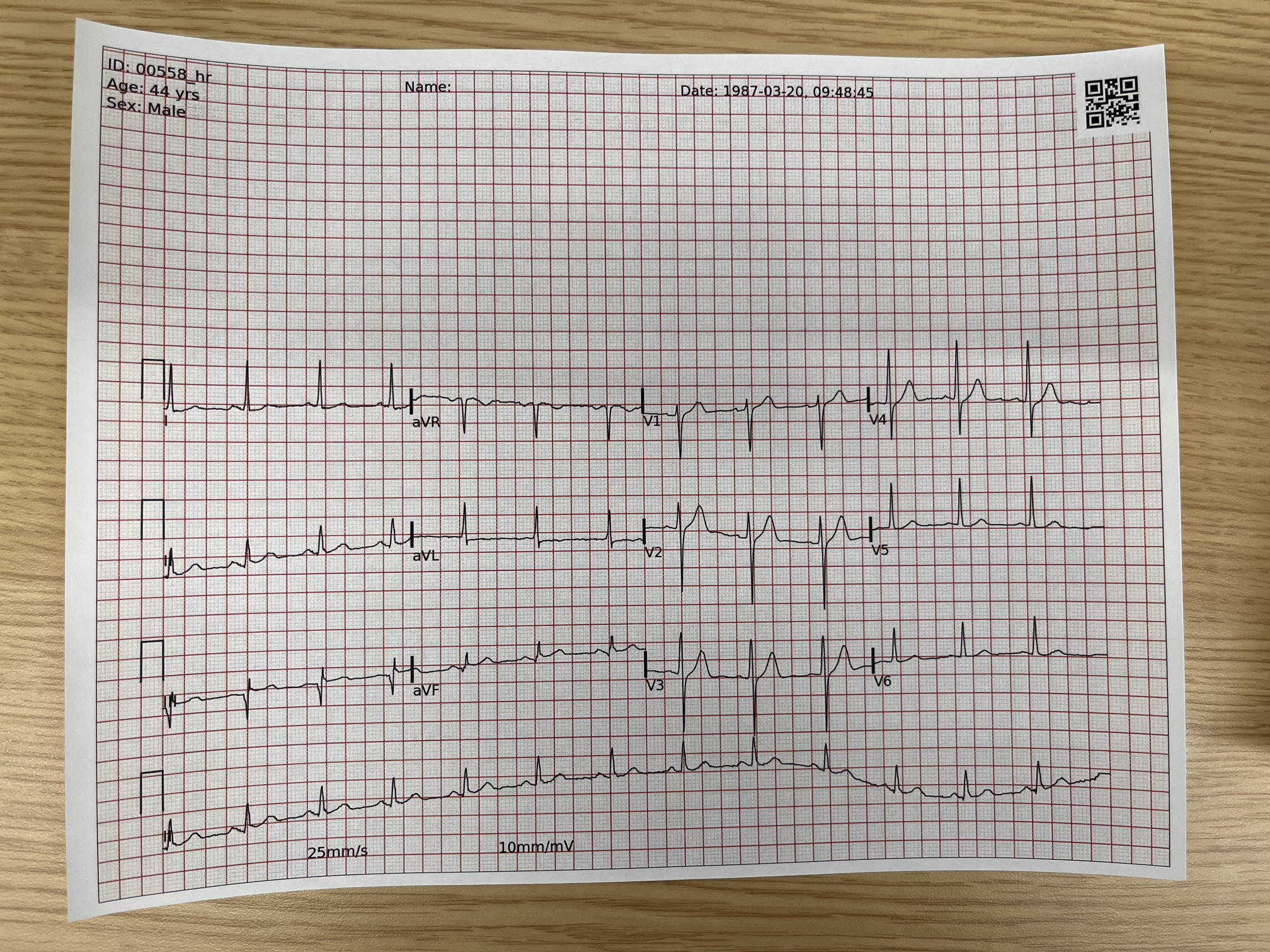}
\caption{Mobile photograph of clean paper (\ref{sec:photo-clean-paper})}
\end{subfigure}
\hfill
\begin{subfigure}[t]{0.325\linewidth}
\includegraphics[width=\linewidth]{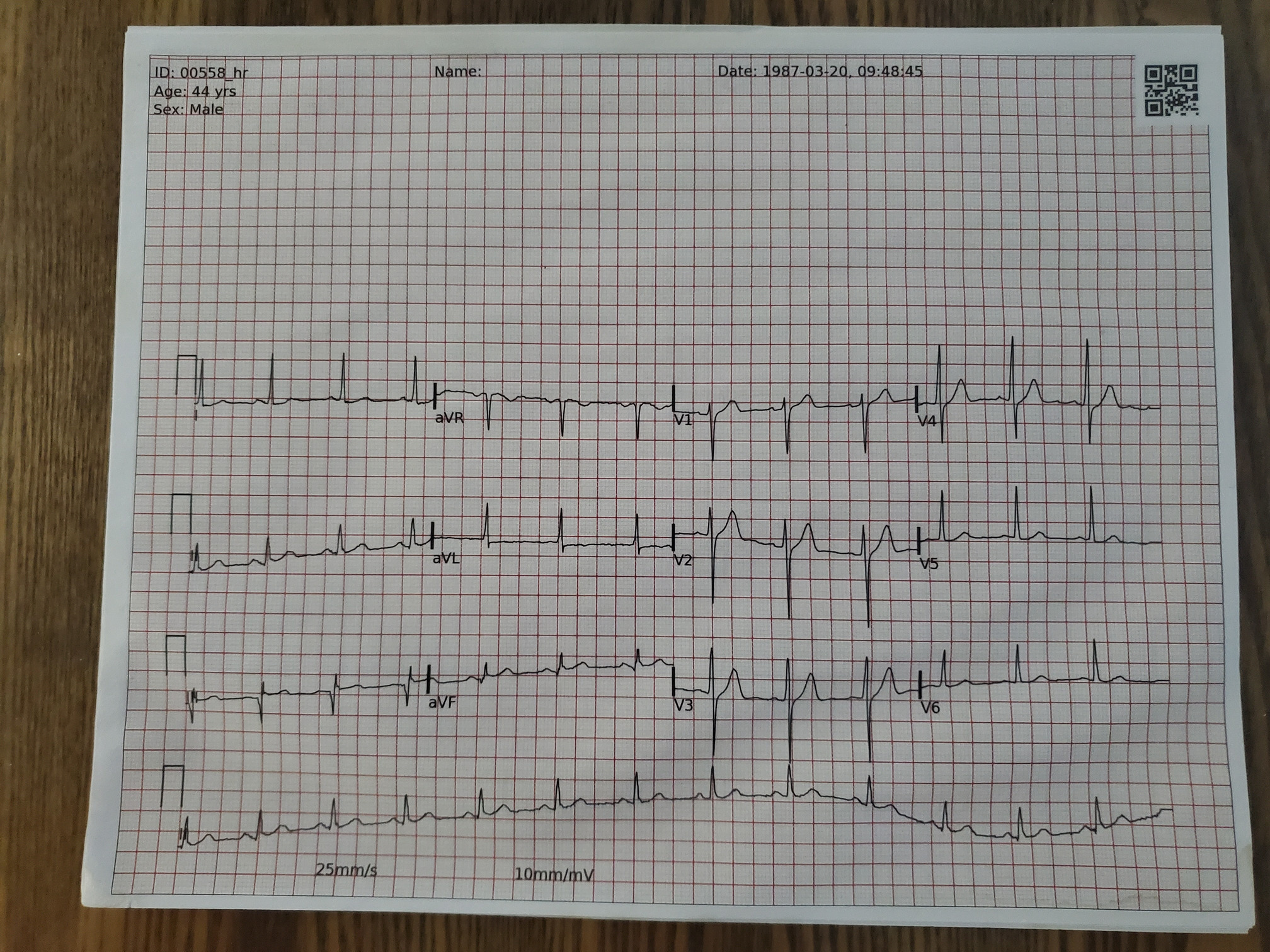}
\caption{Mobile photograph of water-damaged paper (\ref{sec:photo-water-paper})}
\end{subfigure}
\hfill
\begin{subfigure}[t]{0.325\linewidth}
\includegraphics[width=\linewidth]{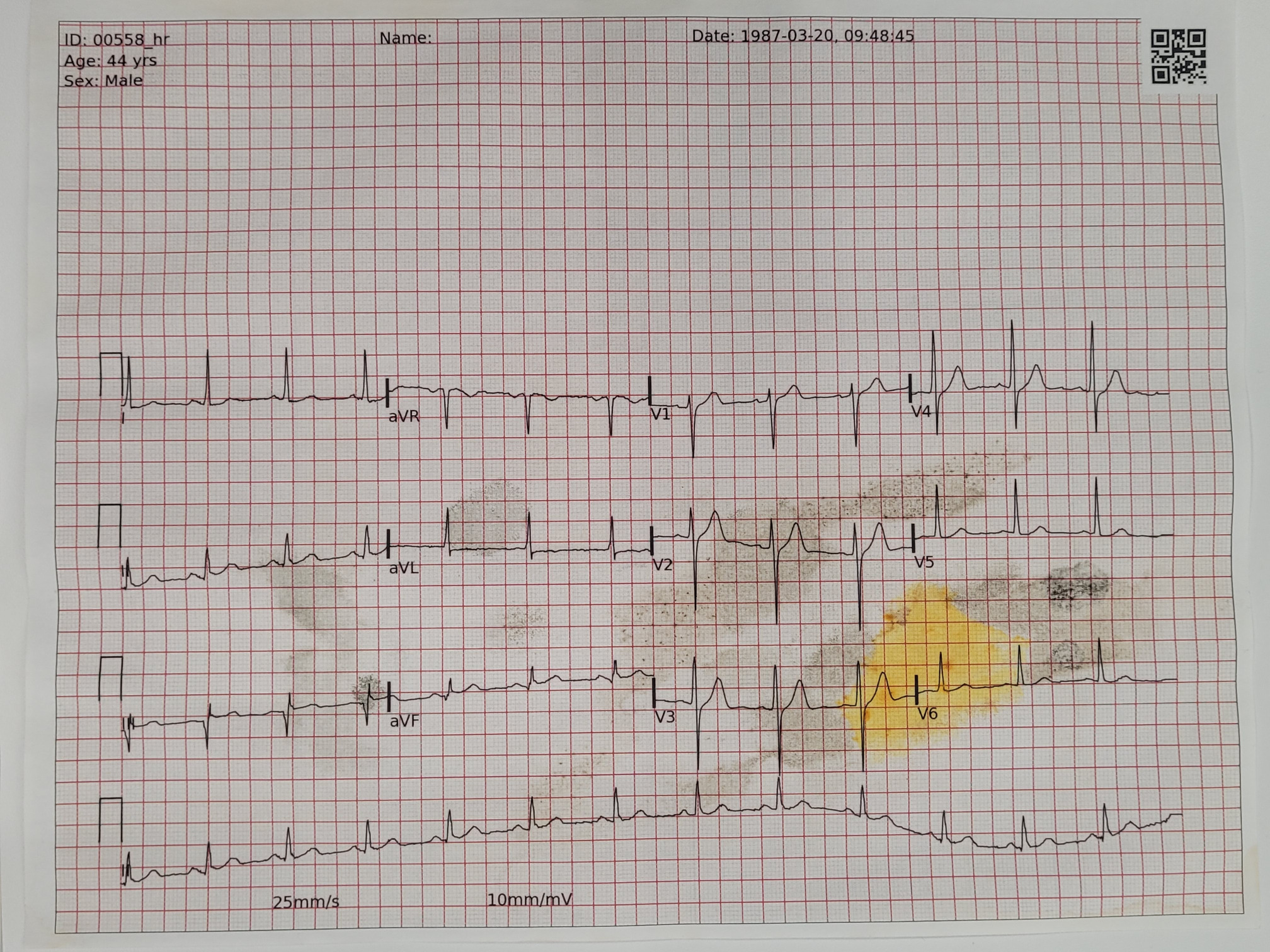}
\caption{Mobile photograph of deteriorated paper (\ref{sec:photo-mold-paper})}
\end{subfigure}
\\
\begin{subfigure}[t]{0.325\linewidth}
\includegraphics[width=\linewidth]{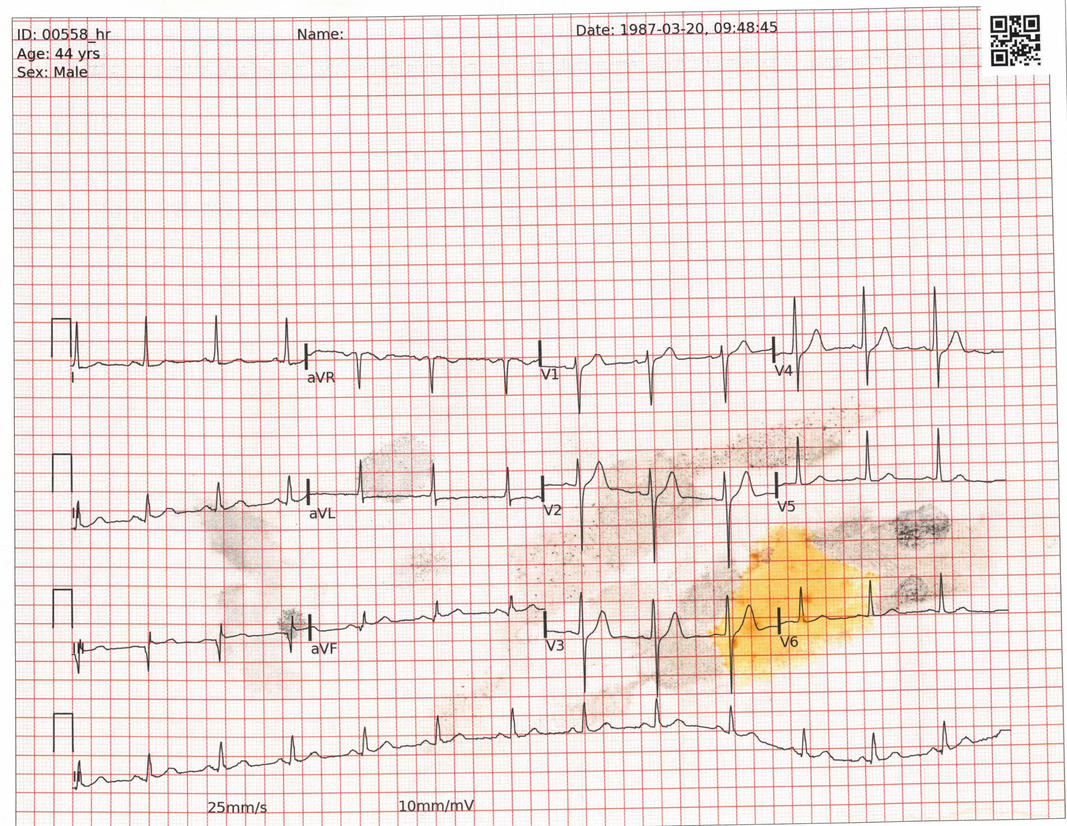}
\caption{Color scan of deteriorated paper (\ref{sec:scan-mold-paper})}
\end{subfigure}
\hfill
\begin{subfigure}[t]{0.325\linewidth}
\includegraphics[width=\linewidth]{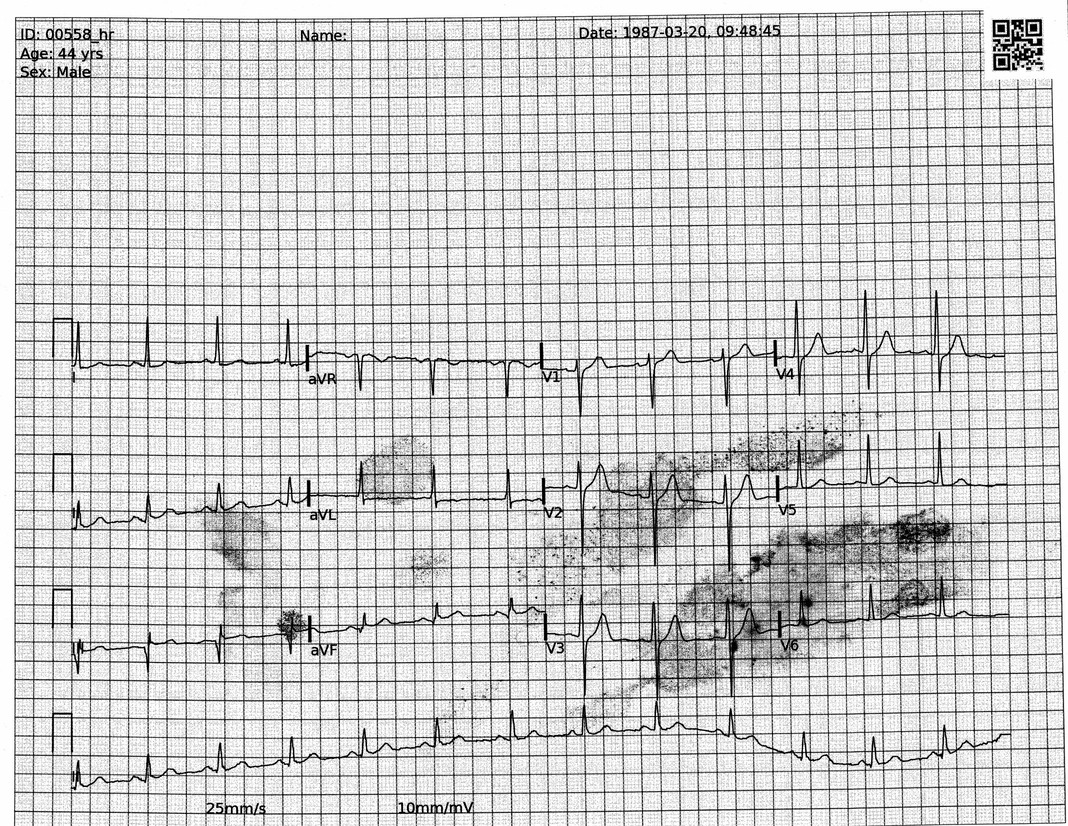}
\caption{Black-and-white scan of deteriorated paper (\ref{sec:scan-mold-paper})}
\end{subfigure}
\hfill
\begin{subfigure}[t]{0.325\linewidth}
\includegraphics[trim={0cm 15cm 0 10cm},clip, width=\linewidth]{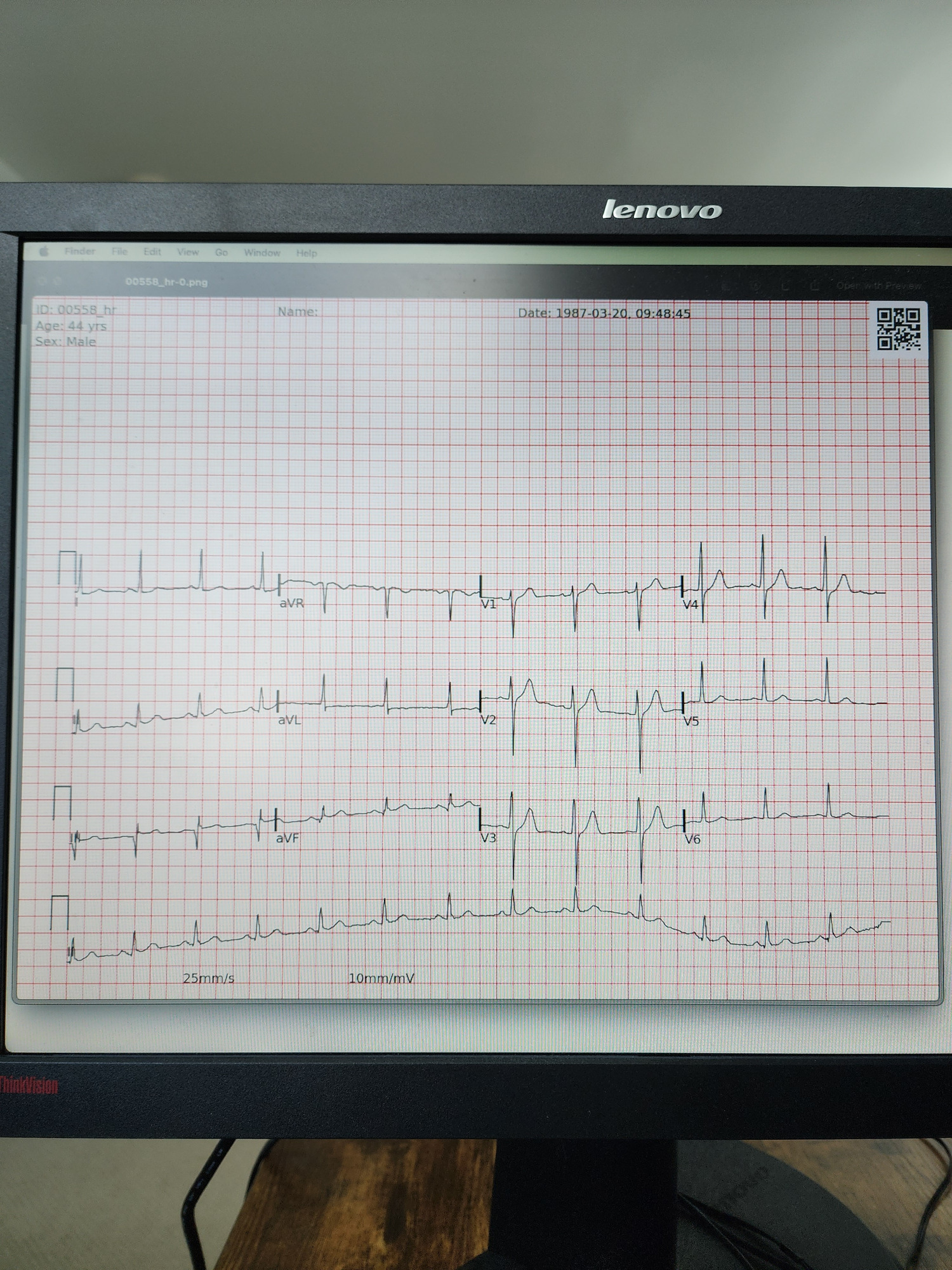}
\caption{Mobile photograph of computer monitor (\ref{sec:photo-monitor})}
\end{subfigure}
\caption{Representative ECG images from the ECG-Image-Database dataset.}
\label{fig:example-figures}
\end{figure}

The PTB-XL and Emory datasets were provided to ECG-Image-Kit, to generate the initial electronic version of the ECG images from time-series data, which were later printed in paper and used to create the various representations of the dataset for ECG image digitization assessment. The different electronic and physical variants of the dataset are detailed below. The different variants of the images created by this procedure, along with their relationships, are illustrated in Figure~\ref{fig:ecg-image-dat-diagram}. Examples of these ECG image variants are shown in Figure~\ref{fig:example-figures}, for a sample record of the PTB-XL dataset. A similar procedure applies to the Emory Healthcare dataset.

\subsection{Dataset naming and version control convention}


ECG-Image-Database hosts various versions of ECG images generated from ECG time-series data. The dataset is version-controlled and will continue to evolve over time. To facilitate the comparison of algorithms trained and developed using this dataset, we have adopted a naming convention that reflects the various challenges in digitizing ECG images and to accommodate future expansions of the dataset. Our naming convention for the ECG image variants follows this format: \texttt{D[m].[n].[p]-v[x]}, where \texttt{m}, \texttt{n}, and \texttt{p} denote the major, minor, and patch levels, respectively, and \texttt{x} denotes the dataset expansion version number.

For example, as detailed below, \texttt{D1.1.0-v1} and \texttt{D1.2.0-v1} represent the first versions of ECG images in .png format generated using ECG-Image-Kit with red and green grids, respectively. The digitization of both datasets is presumed to be at the same level of difficulty for well-designed digitization software. Therefore, they are both prefixed with `\texttt{D1}'. Future expansions in the number of records in each group will be enumerated as \texttt{D1.1.0-v2} and \texttt{D1.2.0-v2}, and so on. Minor modifications or patches (due to potentially erroneous records) will be enumerated as \texttt{D1.2.0-v1}, \texttt{D1.1.1-v1}, and so on. 

Any minor corrections or patches to previously released data will be listed as minor revisions or patches and reported in the release documentation. Updates to the dataset will also be tracked and documented publicly using a DOI for the updated versions, along with notes on the changes. Previous versions of the dataset will remain accessible via their respective DOIs. 





\subsection{Generating ECG images from time-series data}
\subsubsection{Images with red grid.}
\label{sec:images_with_red_grid}
After selecting the ECG time-series records, the first electronic version of the dataset was generated using ECG-Image-Kit. For reproducibility, the subsequent steps for converting the time-series data into images is detailed below.
\begin{enumerate}
    \item \textit{Step 0: Clone the required repositories.} Use your preferred tool/software to clone 
    ECG-Image-Kit: \url{https://github.com/alphanumericslab/ecg-image-kit} and the PhysioNet Challenge 2024 codebase: \url{https://github.com/physionetchallenges/python-example-2024.git},
    \item \textit{Step 1: Downloading the time-series datasets.} In our open-access dataset, the Emory Healthcare time-series data is provided in WFDB format, with annotations included in the WFDB header files. The PTB-XL and PTB-XL+ datasets are not hosted in our repository in time-series format, but they are publicly available on PhysioNet \cite{wagner2020ptb,PTB-XL-PhysioNet,PTB-XL+PhysioNet,Goldberger2000}. PhysioNet provides both interactive and programmatic download options. If downloaded in compressed formats, the data should be unzipped for subsequent processing.
   \item \textit{Step 2: Convert the data to WFDB (WaveForm DataBase) format.} This can be done using the following script from the \cite{2024ChallengePythonCode} for the PTB-XL and PTB-XL+ datasets. The \verb|data| argument is a folder with records from the PTB-XL dataset, and the other files are metadata from the PTB-XL or PTB-XL+ datasets.
\begin{verbatim}
python prepare_ptbxl_data.py \
    -i  data \
    -pd ptbxl_database.csv \
    -pm scp_statements.csv \
    -sd 12sl_statements.csv \
    -sm 12slv23ToSNOMED.csv \
    -o  data
\end{verbatim}
\item \textit{Step 3: Generate synthetic ECG images.} Next, we generate synthetic ECG images from the dataset using the following command from ECG-Image-Kit:
\begin{verbatim}
python gen_ecg_images_from_data_batch.py \
    -i data \
    -o data \
    --add_qr_code \
    --print_header \
    --mask_unplotted_samples \
    --store_config 2
\end{verbatim}
The ECG-Image-Kit command line options selected above and the values for the parameters with  default values are listed in Table~\ref{tab:ecg-image-kit-default-params-red-grid}. More specifically, \verb|gen_ecg_images_from_data_batch.py| uses the default parameters and flags to generate ECG images with red grids in the background. The \verb|--add_qr_code| flag generates QR codes containing the file names in the top-right corner of the generated ECG images. QR codes, or Quick Response codes, are two-dimensional pictographic codes that provide faster/programmatic readability and greater storage capacity than traditional barcodes. In our application, the QR codes facilitate the automatic batch scanning and renaming of the scanned/photographed images, ensuring that the reference time-series and the corresponding ECG images match one another for building machine learning models. For this step, we used the Python module named \verb|qrcode|, which utilizes the Python Imaging Library (PIL) to create QR codes. This module offers both simple and advanced usage options, allowing users to generate QR codes with various levels of customization. The \verb|qrcode| module allows users to control parameters such as version (size), error correction level, box size, and border thickness, enabling the creation of QR codes tailored to specific requirements. Using the qrcode library, we encode the signal filename used to generate the corresponding synthetic image.

\item \textit{Step 4: Add image file locations to WFDB header files.} Add the file locations and other relevant information for the synthetic ECG images to the WFDB header files. This can be accomplished with the following command:
\begin{verbatim}
python prepare_image_data.py \
    -i data \
    -o data
\end{verbatim}
\item \label{prepare_data}\textit{Step 5: Prepare data for training/inference (optional).} To prepare a version of the data for inference (for instance to train models for ECG image digitization or classification), the user will need to remove the time-series waveforms, certain information about the waveforms (like the diagnoses and/or the demographics). This version of the data can be used for the inference step.

\begin{verbatim}
python gen_ecg_images_from_data_batch.py \
    -i data \
    -o hidden_data \
    --add_qr_code \
    --print_header \
    --mask_unplotted_samples

python prepare_image_data.py \
    -i hidden_data \
    -o hidden_data

python remove_hidden_data.py \
    -i hidden_data \
    -o hidden_data \
    --include_images
\end{verbatim}

\begin{table}[tb]
    \caption{Default parameters of \texttt{gen\_ecg\_images\_from\_data\_batch.py} used to generate ECG images with red grids as described in Section~\ref{sec:images_with_red_grid}}
    \centering
    \begin{tabular}{|c|c|c|}
        \hline
        Parameter & Value & Description \\ \hline
        \verb|resolution| & 200 & Image resolution \\ \hline
         \verb|num_columns| & 4 & Number of columns for leads in the ECG image \\ \hline
         \verb|full_mode| & II  & Leads to add as long strips at the bottom of the ECG image \\ \hline         
         \hline
    \end{tabular}
    \label{tab:ecg-image-kit-default-params-red-grid}
\end{table}
\end{enumerate}

\subsubsection{Images with green grid:}
Some ECG device manufacturers use paper with non-red grid colors. ECG-Image-Kit supports arbitrary grid colors through its command line parameters. We used the following command (instead of the command in Step \ref{prepare_data} above) to generate green ECG grid paper, which is common in some ECG machines: 
\begin{verbatim}
python gen_ecg_images_from_data_batch.py \
    -i data \
    -o hidden_data \
    --add_qr_code \
    --print_header \
    --mask_unplotted_samples \
    --standard_grid_color 4
\end{verbatim}
where \texttt{4} is the color code for green.

\subsection{Programmatic distorted variants of the ECG images}
ECG-Image-Kit can be configured to generate realistic ECG image distortions that resemble real-world paper distortions. ECG-Image-Database hosts several variants of the PTB-XL and Emory datasets with these programmatically generated distortions, which are available for reference comparisons in ECG digitization research. Due to the stochastic feature of the toolbox in generating random distortions, similar commands can be used to generate much larger datasets with similar realistic distortions for training data-demanding deep models.

\subsubsection{Creases:}
Creases are simulated by adding evenly-spaced lines to mimic paper fold creases. Gaussian blurring, a common image augmentation technique is applied to these lines to introduce smoothing effects. The blurring enhances realism by creating a shadow-like effect in the creases, common in scanned images or real paper ECG. Wrinkles, on the other hand, are treated as textures, which can be generated using texture synthesis methods like image quilting. To add the wrinkles and creases to the generated images, we use the \verb|--wrinkles| flag and set crease angle (\verb|-ca|) to the desired angle (45 degrees in the example below):
\begin{verbatim}
    python gen_ecg_images_from_data_batch.py \
    -i data \
    -o data \
    --wrinkles \
    -ca 45 \
    -se 10 \
    --random_grid_color \
    --add_qr_code
\end{verbatim}

\subsubsection{Rotations:}
To add rotations to the generated images, we set the augment attribute and set rotation flag (\verb|-rot|) to the desired angles (30 degrees in the example below), the cropping percentage (\verb|-c|) value (0.1 in the example below) and select \verb|--deterministic_rot| and \verb|--deterministic_noise| flags. The deterministic flags ensure a deterministic behavior with no randomness, guaranteeing reproducibility of the process. The full command line is as follows: 
\begin{verbatim}
    python gen_ecg_images_from_data_batch.py \
    -i data \
    -o hidden_data \
    --augment \
    -rot 30 \
    -c 0.1 \
    --deterministic_rot \
    --deterministic_noise \
    -se 10
\end{verbatim}

\subsubsection{Handwritten text}
To add handwritten text to the images, we set the \verb|hw_text| flag and set the number of words to add (\verb|n|) in handwritten style (4 in the example below), the horizontal offset of the added text (\verb|x_offset|) and vertical offset (\verb|y_offset|) in pixels (30 and 20 pixels, respectively, in the example below). 
\begin{verbatim}
    python gen_ecg_images_from_data_batch.py \
    -i ptb-xl/records500/00000 \
    -o ptb-xl/records500_hidden/00000  \
    --hw_text -n 4 \
    --x_offset 30 --y_offset 20 \
    -se 10 --random_grid_color \
    --add_qr_code
\end{verbatim}

\subsection{Physically distorted variants of the ECG images}
The next step was to introduce physical distortions, including imaging, scanning, photography artifacts, and natural deterioration effects to the ECG printouts. The different variants of the resulting datasets are explained below. We used these data as part of the hidden data for the PhysioNet Challenge 2024  \cite{2024Challenge}.

\subsubsection{Scans of red grid printed in color.}
\label{sec:color-scan-clean-paper}
The complete dataset, comprising both the Emory and PTB-XL databases, was printed using a LaserJet Pro printer (model M501) in color mode at a resolution of 600 dpi on US Letter-sized paper. The printed ECG images were subsequently scanned on a Brother scanner model MFC-L8900CDW in color mode, in batches of 50–100 images, and saved as multi-page PDF documents. These PDF files were then programmatically split into individual images. The QReader Python package, powered by the YOLOv8 model, was used to detect and extract QR codes containing individual file names from each separated image. These file names were then matched against the original dataset of waveform filenames to ensure the accuracy of the printing and scanning process, as well as to identify any missing images due to batch scanning failures or human errors. Next, the images were converted into JPEG format using the Python PIL package, applying a quality level of 90\% (compression level of 10\%). Human experts, familiar with ECG signals, performed a visual inspection of the compressed images to ensure that the readability of the signals was maintained. This process reduced the file sizes to approximately 2–3 megabytes per image, while preserving an acceptable level of detail for ECG interpretation and waveform detection.

\subsubsection{Grayscale scans of red grid printed in color.}
\label{sec:bw-scan-clean-paper}
The entire printed dataset in color, including the Emory and PTB-XL databases, was rescanned in grayscale mode using a Brother scanner model MFC-L8900CDW. The same procedure used for the color scans was followed: the images were saved as multi-page PDF files, which were then split into individual images. An automatic QR code detection system was applied to extract the original waveform filenames from the images. The images were subsequently saved in JPEG format with a 90\% quality level, using the original filenames obtained through QR code detection.

\subsubsection{Photographs of red grid with four different cameras.}
\label{sec:photo-clean-paper}
The red grid printed ECG images were photographed using four different mobile phones: Samsung Galaxy S20 Fe (12-megapixels camera, Emory dataset), iPhone 12 (12-megapixels camera, PTB-XL dataset), iPhone 13 Mini (12-megapixels camera, PTB-XL dataset), and Samsung S10+ (12-megapixels camera, PTB-XL dataset). 
The photos were taken in an office setting with ceiling lights. In some images, minor shadows from the cameras, people, or objects were present. These images were all renamed using the QR code detector and saved to JPEG with 90\% quality level. 

\subsubsection{Soaking and staining distortions.}
The color-printed ECG images were deliberately soaked in water and contaminated with coffee stains and soy sauce. This step was performed to replicate the random deterioration of paper ECGs that accidentally occurs in clinical settings with ECG printouts. The impact of water, coffee, and sauce stains varied across the images; some were significantly damaged, while others showed moderate or very minor effects. The contaminated printouts were spread out on a table and left to partially air-dry for approximately 12 hours under a ceiling fan. Random batches that remained significantly wet were ironed to further dry them, resulting in occasional brownish/tan thermal deterioration or scorching effects. Throughout the experiment, the QR code regions of the ECG images (on the top right corner of each image) were retained to facilitate the automatic sorting and naming of the corresponding image files.

\subsubsection{Photographs of stained and damaged ECG papers.}
\label{sec:photo-water-paper}
The contaminated PTB-XL printouts were photographed using a Samsung Galaxy S10+ mobile phone under sunlight, partial shade, and ceiling light. Most of the printouts remained moist or partially wet at this stage. The photos were taken at a distance that allowed the entire paper to fit within the frame of the mobile phone camera. The photographs were captured from different angles and under various lighting conditions (at night, dawn, and early morning) with varying light levels, resulting in shadows and partial shading in some images. There was no specific order or set of instructions for photographing the ECG records, other than capturing a full image of the ECGs. The photographers were given the liberty to take pictures of a quality that would be considered ``acceptable'' by people familiar with ECG data.

\subsubsection{Growing mold and further drying.}
The partially moist ECG papers (after being stained with water, coffee, and sauce) were packed into envelopes, with approximately 100 images in each. The envelopes were sealed, stacked, and stored in a humid place for six weeks. This caused many of the papers to develop small, moderate, or significant amounts of mold, further deteriorating the printouts. In some cases, the mold on the coffee and sauce stains interfered with the ECG waveforms, resembling extreme natural deterioration of paper ECGs over time, as seen in damp historical hospital archives. The level of deterioration varied across the records. Some papers, which were completely dry before packing, did not develop any mold.


\subsubsection{Scans of moldy red grid; black-and-white and color.}
\label{sec:scan-mold-paper}
The moldy, printed ECG images were scanned using a Brother scanner model MFC-L8900CDW. Batches of 10-20 printed images were fed into the scanner to generate multi-page PDFs. Images were first scanned on the color setting at 600 dpi resolution. Images underwent a second round of scanning on the black and white setting at 600 dpi resolution. Due to their fragile nature, the moldy, printed ECG images caused frequent scanner jams and were often creased during the scanning process. Two printed ECGs were destroyed due to paper jamming during the scanning process. These images were re-preinted and contaminated with stain and rescanned. The resulting PDF files were split into separate image files, renamed using the QR code detector, and saved to JPEG with 90\% quality level.

\subsubsection{Photographs of moldy red grid images.}
\label{sec:photo-mold-paper}
The printed, moldy ECG images were photographed using two different mobile phones: an iPhone 11 and a Samsung Galaxy S20 FE, both with 12 megapixels.

\subsubsection{Photographs of ECG on computer monitors with red grid.} 
\label{sec:photo-monitor}
Taking pictures of ECG monitors and sharing them with peers for review or with medical trainees is a common practice in clinical settings. Technologically, monitors refresh their screens at high rates (e.g., 60\,Hz or higher) to provide high-quality images with seamless transitions for the human eye. However, because mobile phones and cameras capture images in very short snapshots (comparable with monitor refresh rates), with exposure times that are not synchronized with the monitor's refresh patterns, photos of monitors are susceptible to unique aliasing and imaging artifacts that affect the color and resolution of the captured images.

To replicate this effect, the Emory and PTB-XL images were displayed on a Lenovo LT2252PWA 22-inch monitor (1680 ${\times}$ 1050) and photographed using a Samsung S22 Ultra mobile phone camera at a resolution of 12M pixels main camera. To simulate diverse real-life conditions, the lighting, the monitor’s angle relative to the external light source, and the phone’s orientation (landscape or portrait, each with slight variations in angle) were arbitrarily varied by the photographer. The camera distance from the monitor varied between 10 and 15 inches, depending primarily on the orientation. Some blurring was also introduced in some photographs due to hand motion artifacts.

\section{Discussion}
The ECG-Image-Database provides a valuable contribution to the ongoing effort to digitize and analyze non-digital ECG data, which was the focus of the PhysioNet 2024 Challenge on ECG image digitization and classification \cite{2024Challenge}. With the increasing availability of digital ECGs from modern clinical devices, there remains a vast amount of historical ECG data that exists solely in paper or scanned image form, or ECGs in electronic health records for which the underlying ECG time-series has not been stored or maintained. These legacy records contain vital information about patient health histories, rare cardiovascular events, and demographic-specific data that are irreplaceable for longitudinal studies and population-wide analyses. Without proper digitization and analysis tools, much of this valuable information could be lost due to physical deterioration or obsolescence of archival systems.

The development of ECG-Image-Database is aimed at addressing several key challenges in the field of ECG digitization. First, by generating a large collection of high-fidelity synthetic ECG images paired with the underlying ECG images and realistic physical and electronic distortions, the dataset simulates the types of real-world artifacts commonly encountered in clinical practice. This includes noise, wrinkles, stains, perspective shifts, and other degradations that often complicate the digitization process. In addition to digitally induced distortions, the inclusion of physically altered images---through soaking, staining, and mold exposure---provides a comprehensive collection of ECG images that mirror those found in real-world clinical archives. This level of realism helps to ensure that machine and deep learning models trained on this dataset are able to generalize well to diverse input conditions, an important consideration for widespread adoption in clinical settings.

Moreover, the dataset's structure, which currently includes 35,595 ECG images generated from 1,977 unique time-series records, offers a rich resource for developing robust and generalizable algorithms. By providing paired time-series and image data, ECG-Image-Database enables researchers to explore different modalities of ECG analysis. This could result in the development of models that not only perform digitization (converting image data back into time-series signals which can be used for AI-ML applications) but also directly classify conditions based on image data alone, bypassing the need for digitization altogether. Such dual approaches could lead to the development of more efficient tools for ECG analysis using the advantages of ECG as a time-series and as an image. As a matter of debate, clinicians are more familiar with the latter holistic approach (looking at all ECG leads at a glance), while traditional ML pipelines are mostly based on the former.

Another important contribution of this dataset is its role in preserving cardiovascular data from non-digital formats. In many low-resource regions and low-to-middle-income countries (LMICs), digital ECG infrastructure remains scarce, and healthcare providers often rely on printed ECGs. Even in high-resource settings, paper ECGs are still used in educational settings, clinical decision-making, and peer discussions. The ECG-Image-Database can serve as a critical reference for the development of accessible, low-cost digitization tools that can be deployed globally. This opens the door for equitable access to modern cardiovascular diagnostic tools, reducing disparities in health outcomes across different regions and populations.

The application of this dataset in the 2024 PhysioNet Challenge further underscored its importance. The challenge aimed to synergize innovation and efforts in ECG digitization and classification, encouraging researchers to develop models capable of handling real-world ECG data with varying levels of degradation. Early results from the challenge suggest that combinations of conventional signal processing, computer vision, and deep learning models are capable of extracting time-series data from ECG images with promising accuracy, but there is still significant room for improvement, particularly when dealing with extreme distortions or artifacts.

Despite these advances, several challenges remain. One of the key hurdles in ECG digitization is dealing with low-quality images, especially those captured through suboptimal conditions such as poor lighting, incorrect angles, or damaged physical ECGs. While the ECG-Image-Database provides a range of such examples, real-world variability is difficult to fully capture in a controlled setting. Another challenge is ECG images with protected health information (PHI). While ECG-Image-Database provides tools for generating printed and handwritten-style text on the generated ECG images, the dataset does not currently host examples of these cases, which could be used to train ML algorithms to remove PHI, and/or to evaluate their robustness to PHI leakage or any adversarial attacks. 

Another aspect requiring further exploration is the evaluation of models for ECG image digitization. The 2024 PhysioNet Challenge employed various signal-to-noise ratio (SNR) metrics to measure the quality of ECG image reconstruction by the participants, which was feasible because the ECG-Image-Database provides the ground-truth time-series data. However, for historical ECG images where we do not have access to the ground-truth, applying SNR metrics is unfeasible. Moreover, SNR is not necessarily the best metric to assess the fidelity of a digitization algorithm in preserving ECG details that are significant for ECG-based diagnosis. While SNR captures major discrepancies---such as distortion in (or loss of) the QRS complex or T-wave---it might not fully reflect the clinical significance of more subtle and low-power details of the ECG, such as the Q-wave onset/T-offset or the ST segment, which are crucial for accurate diagnosis. In future research, ECG-specific SNR metrics (e.g., weighted SNR measures) could be used to emphasize diagnostically important components of the recovered ECG. Alternatively, as highlighted in the 2024 PhysioNet Challenge, the focus might shift away from time-series recovery and toward the \textit{diagnostic value} of the recovered ECG. Hypothetically, an ECG image digitization scheme that preserves the ``essential diagnostic'' elements of the ECG could be preferable, even if it does not achieve the highest SNR. However, when SNR values reach higher values (approximately 20--30\,dB or higher), which corresponds to what human experts would visually consider an excellent reconstruction, the reconstruction would generally be deemed acceptable for building ECG diagnosis ML models.

Finally, although the ECG-Image-Database fills a critical gap in the availability of paired ECG time-series and image data, the field would benefit from continued expansion of this resource. We have adopted a file naming convention that would facilitate future expansions, in the number of underlying ECG time-series and various physical/electronic distortions. Future versions could include additional ECG devices, different printing technologies, and more varied environmental distortions to further increase the robustness of models trained on this dataset. Additionally, while the dataset currently focuses on 12-lead ECGs, expanding to other types of ECG configurations (such as single-lead or 3-lead recordings) could broaden its applicability to applications like wearable and portable ECG devices.

\section{Conclusion}
The ECG-Image-Database offers a standardized resource for developing machine and deep learning models to digitize and classify ECG images, by reproducing real-world challenges such as image distortions, noises, and environmental artifacts. By replicating the conditions under which ECGs are stored and scanned, the dataset ensures models trained on it can effectively handle real-world complexities, preserving valuable diagnostic information from paper-based ECG records.

Future expansions, including new image sources and further environmental variations, will enhance the dataset's utility for training models that generalize across diverse conditions, supporting both clinical and low-resource applications in cardiovascular diagnostics.

\section*{Acknowledgements}
G.D.~Clifford, M.A.~Reyna and R.~Sameni are supported by the National Institute of Biomedical Imaging and Bioengineering (NIBIB) under NIH grant number R01EB030362. Ethical approval for this work was granted by Emory's Institutional Review Board (Ref: STUDY00007353). 

\appendix
\section{A note on ECG image vs time-series resolutions}

\subsection{ECG as a time-series}
The ECG recorded by standard body surface leads typically has an amplitude of several millivolts, and its spectral content ranges from approximately 0.05\,Hz to around 150\,Hz. When transformed into a digital signal, it first passes through an anti-aliasing low-pass filter (in the analog domain) before being sampled at a sampling frequency $f_s$. According to the Nyquist theorem, the cutoff frequency of the anti-aliasing filter should be lower than $f_s/2$. Systems with lower resolution, such as old single-lead ECG machines, Holder monitors and wearable devices, may have sampling frequencies as low as 100\,Hz. Modern high-quality clinical monitors often sample at higher frequencies, such as 1000\,Hz or higher.

The amplitude resolution of the digital signal depends on the number of bits of the analog-to-digital converter (ADC), which we denote by $N$. In older ECG devices, $N$ was as low as 8 bits, resulting in a maximum resolution of 256 quantization levels (assuming that the ECG was amplified to span the ADC's full dynamic range). In modern ECG devices, $N$ can be up to 24 bits, resulting in significantly better amplitude resolution. Importantly, due to electronic and thermal noise, and depending on the quality of the analog front-end circuitry, the effective number of bits (ENOB) is, in practice, lower than the nominal ADC bit number $N$. For instance, a 16-bit ADC may yield between 12.5 to 14 ENOBs in practice. With an $N$-bit ADC digitizing the input voltage range of $V_{\min}$ to $V_{\max}$, the voltage resolution after digitization is:
\begin{equation*}
\delta v = \frac{V_{\max} - V_{\min}}{2^N}
\end{equation*}
For example, if the input span of the ADC is $\Delta V=V_{max} - V_{min} = 5mV$, an 8-bit ADC yields a voltage resolution of $\delta v=19.5\mu V$, while a 12-bit ADC yield $\delta v=1.22\mu V$, improving the voltage resolution by a factor of 16.

\subsection{Printed ECG}
In clinical applications, the ECG is printed on standard ECG paper featuring fine grids and coarse grids. The fine grids are 1\,mm by 1\,mm, corresponding to 0.1\,mV in amplitude and 40\,ms in time. The coarse grids are 5\,mm by 5\,mm, corresponding to 0.5\,mV in amplitude and 200\,ms in time, as shown in the image below.

\begin{figure}[tb]
    \centering
\includegraphics[width=0.4\textwidth]{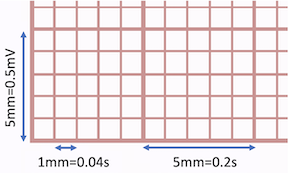}
    \caption{Standard grid lines printed on both hardcopy printouts and digital ECGs and used by clinicians (and algorithms) as references for measuring time and amplitude units of an ECG.}
\end{figure}

In modern ECG devices, despite the data being collected digitally and stored on computers or other digital platforms, the same convention is used. ECGs are visualized against the same background grids, whether displayed on a computer screen or printed as a PDF or image file. The number and format of the leads, along with the ECG grid color, vary depending on the ECG acquisition technology and the device manufacturer. The most common clinical ECGs are 12-lead, featuring approximately 2.5\,s segments of the 12 leads arranged in a 3-row by 4-column grid. Additionally, one to three leads (typically leads II, V1, V2, or V5) are displayed as a longer 10-second strip at the bottom, as shown in the examples throughout the paper.

\subsection{Scanned ECG images}
Printing an analog or digital ECG on paper and then rescanning it as an image involves implicit or explicit interpolation and resampling of the original ECG. When performed by an analog machine or a standard printer, it involves digital-to-analog circuitry to convert the discrete time samples into a continuous waveform, printed as a continuous curve on the paper. Once an ECG is printed, the original sampling frequency $f_s$ of the digital time series, and the number of its quantization bits $N$, become irrelevant, as the signal has been transformed back into the continuous-time domain. When the ECG paper is scanned or photographed as an image, it is essentially being quantized and resampled again, this time as a two-dimensional image. Assuming the image is scanned at a resolution of $D$ dots per inch (DPI), each 1-inch by 1-inch square of the printed ECG is quantized into a $D \times D$ array, each pixel stored in $B$ bits (this is different from $N$, the number of ADC bits). If scanned as a color image, there will be three such arrays, corresponding to the colors red, green, and blue. Modern images typically use $B = 8$, resulting in 24 bits, or 3 bytes, per pixel. Therefore, for example, scanning a letter-size paper of 11 inches by 8.5 inches at 72 DPI would require 8.5 $\times$ 11 $\times$ (72 $\times$ 72) $\times$ 3 bytes = 1,454,112 bytes, or 1.39\,MB, as an uncompressed bitmap file (excluding any metadata or other headers stored in the file). In practice, fewer bits, corresponding to a lower color depth, and/or lossy or lossless image compression, can reduce the storage needed to save such an image with minimal or no loss of information, although scanning and other artifacts may impede compression.

Therefore, when a standard ECG, printed on A4 or letter-size paper, is scanned at full image size (without any cropping or excess borders), each 1\,inch (25.4\,mm) horizontally and vertically maps to $D$ pixels. In other words, each coarse square of the ECG (0.5\,mV in amplitude and 200\,ms in time) maps to a square of $\left(\frac{5 \times D}{25.4}\right)\times\left(\frac{5 \times D}{25.4}\right)$ pixels. This means the amplitude resolution of the scanned ECG is:
\begin{equation}
dv = \frac{2.54 mV}{D}
\end{equation}
and the temporal resolution is $dt = \frac{1.016s}{D}$ seconds, or equivalently, the sampling frequency (in Hertz) of the ECG (as an image) is:
\begin{equation}
f_s' = \frac{D}{1.016}
\label{eq:image_sampling_freq}
\end{equation}
As we can see, the effective sampling frequency $f_s'$ of the scanned ECG is independent of the original digital signal's sampling frequency $f_s$. However, since the original signal's frequency range was limited to $f_s/2$ by the analog front-end anti-aliasing filter, increasing $D$, and therefore $f_s'$, will yield smoother waveforms, but it will not add any information beyond $f_s/2$ (the Nyquist rate of the original digital ECG time-series).

From (\ref{eq:image_sampling_freq}), it is evident that typical image resolutions, such as 72 or 96 DPI, which are common in image analysis applications, are quite low for ECG scanning and digitization applications, as they only provide sampling frequencies of 70.9\,Hz and 94.5\,Hz, respectively. An ECG should be captured at least 125\,Hz, although 250\,Hz, or even 500\,Hz, is preferable in adults. Based on this analysis, a resolution of at least 150 DPI or higher, without any lossy compression, is recommended for ECG scanning and digitization purposes.

Importantly, the calculation of the ECG grid size from the image DPI and paper size is accurate only when using a standard full-paper size scanner. For ECG images captured by cameras, smartphones, screenshots, or through cropping and resizing, the equivalency of 1 inch on the actual paper to the captured image DPI may not hold true. Consequently, ECG digitization algorithms should estimate the correct grid sizes by employing algorithms that detect and analyze the ECG grid sizes directly from the ECG image. Several functions for this purpose are provided in ECG-Image-Kit \cite{ecg-image-kit-software}. Further details and examples can be followed from \cite{Shivashankara2024}.
\section*{References}
\bibliographystyle{IEEEtran}
\bibliography{egbib}

\end{document}